# Quantum Public Key Distribution using Randomized Glauber States

Randy Kuang
Quantropi Inc.
Ottawa, Canada
randy.kuang@quantropi.com

Nicolas Bettenburg
Quantropi Inc.
Ottawa, Canada
nicolas.bettenburg.@quantropi.com

*Abstract*— State-of-the-art Quantum Key Distribution (QKD) is based on the uncertainty principle of qubits on quantum measurements and is theoretically proven to be unconditionally secure. Over the past three decades, QKD has been explored with single photons as the information carrier. More recently, attention has shifted towards using weak coherent laser pulses as the information carrier. In this paper, we propose a novel quantum key distribution mechanism over a pure optical channel using randomized Glauber states. The proposed mechanism closely resembles a quantum mechanical implementation of the public key envelope idea. The core idea can be described in five steps as follows:

1. A user (Bob) generates a Glauber state as a quantum public key envelope (QPKE) by randomly modulating a secret phase $\varphi_r$, known only to Bob, and transmits it over an optical channel to the other user (Alice).
2. Alice modulates a key phase $\varphi_k$ into the QPKE based on a random key and selected modulation scheme and returns it to Bob.
3. For the returning QPKE, Bob derandomizes it with his private key or the phase $-\varphi_r$ and then
4. passes it to a coherent receiver to measure the key phase $\varphi_k$.
5. For better security, differential phase-shift keying (DPSK) technique with a reference list is applied to extract keys.

For the proposed solution, we explore physical countermeasures to provide path authentication and to avoid man-in-the-middle attacks. Other attack vectors can also be effectively mitigated by leveraging the QPKE, the uncertainty principle and the DPSK modulation technique.

## I. INTRODUCTION

In 1984, Bennett and Brassard proposed a groundbreaking concept titled "Quantum Cryptography: Public Key Distribution and coin tossing" [1], later referred to as quantum key distribution (QKD) or more commonly referred to as "BB84". Shor and Preskill (2000) [2] provided us later with a proof that BB84 offers unconditionally secure key distribution. Since then, QKD has been widely researched with a body of work exploring a variety of implementations as well as improvements on imperfections of physical devices. Diamanti, Lo, Qi and Yuan (2016) [3] reviewed the practical challenges with a variety of QKD practical implementations. More recently, Xu et al (2019) [4] made a more complete review of its security analysis over the protocols, implementations, signal sources, and detections. This body of work is comprehensive, and we would like to point the interested reader to an extended literature review for additional background material [4].

QKD was first proposed to use single photons (Fock states $|n=1\rangle$) as the information carrier, also called "qubit", and this implementation is commonly referred to as "discrete variable QKD" or "DV-QKD". More recently, attention has turned to continuous variable QKD (CV-QKD) [5], with weak coherent states acting as the information carrier. The motivation for such interest lies in the practical requirements of existing network infrastructure, as well as high detection efficiencies with matured detection techniques of coherent states, such as homodyne and intradyne detections [6]. In theory, QKD has been shown to be secure for key distributions, but imperfections entailed by its practical implementation leave doors open to potential attacks [3, 4, 7, 8, 9]. Research to date has primarily focused on how to improve equipment, as well as its setup, in order to reduce imperfections in hardware and signal processing while maintaining the security properties that make QKD so attractive. Little attention has been paid to the protocol itself, i.e., the post-processing over a public channel:

1. Publicly announcing the measuring basis not only leaves a door for photon number split (PNS) [3] attacking but also opens a door for potential man-in-the-middle (MITM) attacks in both classical and quantum channels. An extra pre-shared secret is required for authentication to avoid MITM attacks. More strictly speaking, this pre-shared secret puts QKD into a hybrid solution of quantum and classical.
2. To avoid PNS attacking, ideal single photons or pure Fock states $|n=1\rangle$ must be used for quantum channel. But a controllable single photon source is realistically impractical. The practical choice is to use extremely weak laser pulses which are not Fock states but weak coherent states. That, however, weakens the condition of theoretical security of QKD based on the laws of quantum physics.

For the approach presented in this paper, we take a step back and look for another possible implementation of a QKD system, while at the same time eliminating some of the concerns with the BB84 protocol. More specifically speaking, we take an inspiration from a typical classical public key exchange [10] process.

1. Bob generates a key pair (public key $K_1$, private key $K_2$) and sends $K_1$ to Alice.

2. Alice encrypts her secret s with $K_1$: c = enc(s, $K_1$) and sends the cipher c back to Bob.
3. Bob decrypts the secret s from the cipher c with his private key $K_2$: s = dec(c, $K_2$).

In the above key exchange process, Bob's public key behaves like an envelope for Alice to drop her secret into, Bob can then extract it by using his private key after he receives the cipher back from Alice. The method proposed in our work aims to mirror this behavior with a quantum public key envelope (QPKE), mimicking the classical public key envelope, in order to achieve the task of quantum secure key distribution. Specifically, we propose the use of a randomized Glauber state [11] as a candidate for a QPKE information carrier.

In the remainder of this paper, we will first introduce Glauber states (section 2), then describe the novel proposal (section 3), and finally draw our conclusion at the end.

## II. GLAUBER STATES AND COHERENT DETECTIONS

In quantum mechanics, coherent states are the special quantum states of the quantum harmonic oscillator with its energy eigenstate |n⟩ obtained by solving its Schrödinger equation of its Hamiltonian $\hat{H}$. They were first introduced to the quantum theory of light by R. J. Glauber in 1963 [11], and later are referred to as Glauber states. Glauber states are very important in today's high-speed optical communication infrastructure, especially in long-haul, metro and wireless backhaul networks. A Glauber state is usually denoted by a Dirac bra-ket notation with α to be |α⟩. It is an eigenstate of the annihilation operator $\hat{a}$ of the quantum harmonic oscillator system:

$$\hat{a} |\alpha\rangle = \alpha |\alpha\rangle, \quad (1)$$

and the annihilation operator $\hat{a}$ in Eq. (1) is not Hermitian, so it has a complex eigenvalue $\alpha = |\alpha| e^{i\varphi}$ with |α| as its amplitude and φ as its phase. The relationship between the Glauber state |α⟩ and system energy eigenstates |n⟩ is that |α⟩ is a superposition state of energy eigenstates |n⟩, also called Fock states (used in DV-QKD), with a probability amplitude associated with the Poissonian number distribution. Therefore, a Glauber state has an indefinite number of photons and a precisely defined phase, which makes a Glauber state an attractive information carrier through possible amplitude and phase modulations.

A quantum harmonic oscillator has two sets of conjugate variables: photon number n with phase φ and coordinate or in-phase q with momentum or quadrature p. Their uncertainty relationships are:

$$\Delta p \, \Delta q \geq 1 \quad (2a)$$
$$\Delta n \, \Delta \varphi \geq \tfrac{1}{2} \quad (2b)$$

The above uncertainty relationships in Eqs. (2a) and (2b) play a critical role in coherent detections. Beck, Smithey and Raymer had experimentally verified the number-phase uncertainty relationship in Eq. (2b) [12]. Their experiment demonstrates that the number-phase uncertainty maximizes at around single photon coherent states where $\Delta n \Delta \varphi \approx 3/4$ and then tends to ½ as the average photon number increases beyond 4 photons per pulse. Figure 1 is cited from their paper

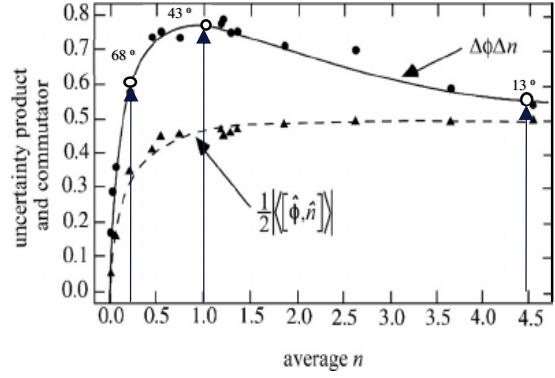

Figure 1. Measurement of the number-phase relationship of the optical field [12]. The top solid curve denotes the number-phase uncertainty and the bottom dashed curve shows the commutator. The numbers besides open circles indicate the phase uncertainties at the corresponding average photon numbers

by inserting three typical points, marked with open circles: $\bar{n}$ = 0.25, $\bar{n}$ = 1.0 and $\bar{n}$ = 4.5. The corresponding phase uncertainties are shown beside the circles. For $\bar{n}$ = 0.25, which is a typical intensity in DV-QKD, the phase uncertainty Δφ is about 68 degree; at $\bar{n}$ = 1.0, the uncertainty reduces to 43 degree and then down to 13 degree when the average photon number increases to 4.5. One can easily extend the curve to higher average photon numbers, the phase uncertainty slowly decreases to 3 degree at $\bar{n}$ = 100.

A Glauber state in phase space is illustrated in Fig. 2. Inside this figure, the small round circles indicate the uncertainty of Glauber states. This uncertainty circle is the same for different Glauber states from the ground state where $|\alpha=0 \, e^{i\varphi}\rangle$ is the quantum noise with an energy $E_0 = \tfrac{1}{2} \hbar\omega$ (ℏ is plank constant and ω is the circular frequency of the

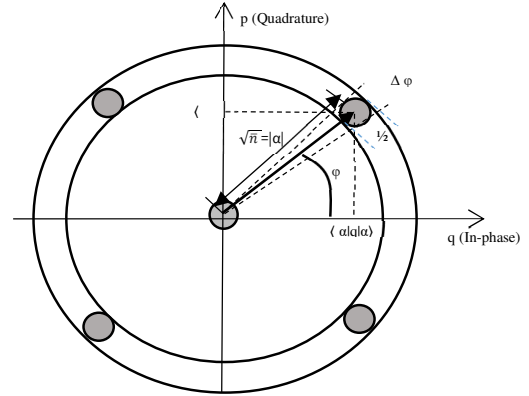

Figure 2. Glauber state is plotted in a phase space. Coordinates are denoted by q as in-phase and p as quadrature. The eigenvalue α has an amplitude |α| and a phase φ. Δφ indicates the uncertainty of a phase. Circles denote the quantum noise for different Glauber states.

oscillator), to any excited state with large amplitude |α|. When the amplitude |α| is well less than 1 (this is where DV-QKD usually works and see the experimental measurement in Figure 1) the quantum signals are within or at the quantum noise level. In this area, the quantum noise plays a critical role

in the measurement of qubits, which is the major factor of lower key rate and relative short distance for key distributions. On the other hand, at very large $|\alpha|$, the quantum harmonic oscillator tends to resemble a classical harmonic oscillator in which the quantum noise become less critical. For optical communications, the intensity of laser pulses is selected to be a suitable level by considering the modulation scheme and the transmission distance. To overcome the significant difficulties associated with DV-QKD, a major shift in information carrier was made, from single photons to weak coherent states with Gaussian modulations for signals, or CV-QKD [5]. CV-QKD is also based on the advances from the matured coherent detections. Coherent detections have two inputs; one is the signal, and the other, a local oscillator LO. The strong LO can boost the signal intensity by $\sqrt{I}$ times, where I is the LO intensity. Although the information carrier is changed to weak coherent states or Glauber states, CV-QKD still adapts the same QKD protocol with both the classical channel for post-processing and the quantum channel for key distributions. In fact, we are more interested in an exploration of a potential change of the BB84 protocol to take the advantage of the changed information carrier in CV-QKD while still offering secure key distribution over today's deployed telecommunication optical infrastructure at a key rate of optical communication speeds.

In the following, we describe our proposed quantum public key mechanism by using a self-control randomized Glauber state as a public key envelope. The proposed approach employs industry-standard modulation schemes and coherent detections, such that it can be readily implemented in existing fiber-optic communication hardware. In addition, we provide special consideration of security properties within the context of the uncertainty principle for Glauber states.

### III. PROPOSED QUANTUM PUBLIC KEY DISTRIBUTION

An overview of our proposed approach is presented in Fig. 3. Laser diode LD generates coherent laser pulses or Glauber states; the phase modulator PM in randomizing module RM takes a random number from random number generator RNG and maps it to a voltage to drive the phase modulator PM, produces a QPKE or a Glauber state $|\alpha_r\rangle$ with a corresponding random phase $\varphi_r$ and transmits the QPKE over an optical channel to Alice. Alice receives the incoming QPKE; her key modulator KM takes a random key from random number generator RNG and performs phase and amplitude modulation to produce a cipher QPKE, with a key phase $\varphi_k$ and an amplitude $|\alpha'_r|$ from the key-modulation mapping table; the cipher QPKE $|\alpha'_r\rangle$ is then returned to Bob. Bob derandomizes the cipher QPKE with his $PM^\dagger$ to remove the random phase $\varphi_r$; the QPKE is detected by Glauber state detector $\widehat{GSD}$ to extract the key phase $\varphi_k$ and amplitude $|\alpha'_r|$ modulated by Alice.

The roundtrip mechanism gives an advantage to Bob that he does not need to tell Alice which random phase was modulated to the QPKE and Alice does not need to know it either to modulate the QPKE with her secret s. Alice just treats it like a normal Glauber state as information carrier.

When the cipher QPKE returns to Bob, the phase is

$$\Phi_k = \varphi'_k + \varphi_r \qquad (3)$$

where $\varphi'_k$ in Eq. (3) is an overall phase except for Bob's random phase $\varphi_r$ and the phase modulator $PM^\dagger$ plays the role of a private key to remove the random phase $\varphi_r$ from Eq. (3):

$$\Phi'_k = \Phi_k - \varphi_r = \varphi_k + \varphi_p + \varphi_{LO}, \qquad (4)$$

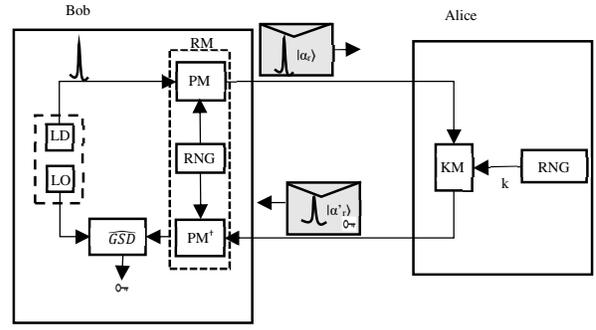

Figure 3. The proposal is illustrated. LD: laser diode, LO: local oscillator, RM: randomizing module, PM: phase modulator, $PM^\dagger$: reverse PM, RNG: random number generator, KM: key modulator (phase and amplitude), $\widehat{GSD}$: Glauber state detector-usually coherent detector with balanced 2x4 hybrid 90° coupler (LO and the returning Glauber state are its two inputs).

Eq. (4) shows the phase before the detector $\widehat{GSD}$ which consists of

- $\varphi_k$ : the actual key phase modulated by Alice to be extracted.
- $\varphi_p$: an additional phase shift from the whole light path including the impact from its environment. This phase is slowly changed from one pulse to another as time goes in comparison with the baud rate of the laser pulses.
- $\varphi_{LO}$: the relative phase between a signal laser pulse at LD and the local oscillator at LO. LD and LO can be the same laser source and can also be different laser sources. However, the difference between them is also relatively stable from one pulse to another.

To extract the key phase, matured techniques such as constellation diagram and the digital signal process (DSP) can be applied to calibrate and compensate the QPKE signals.

In addition to the above, the differential phase-shift keying or DPSK technique can help to cancel out the slow variant phases $\varphi_p$ and $\varphi_{LO}$. Let's consider a reference state $|\alpha_R\rangle$, Alice modulates a $\varphi_R$ and Bob detects a reference phase with his detector $\widehat{GSD}$ seen from Eq. (4)

$$\Phi'_R = \varphi_R + \varphi_p + \varphi_{LO}, \qquad (5)$$

with the factors $\varphi_p$ and $\varphi_{LO}$ for the reference QPKE. Then using Eq. (4) – Eq. (5), we obtain a differential phase

$$\Delta\Phi_k = \Phi'_k - \Phi'_R = \varphi_k - \varphi_R = \Delta\varphi_k \qquad (6)$$

Eq. (6) shows that DPSK makes a key phase extraction independent from impacts of the light path and the local oscillator. Then Alice and Bob use the exact same method to modulate a key based on $\Delta\varphi_k = \varphi_k - \varphi_R$ so a modulated QPKE state with a phase $\varphi_k = \Delta\varphi_k + \varphi_R$ and extract the key based on $\Delta\Phi_k = \Phi'_k - \Phi'_R$. It is a very attractive modulation scheme for quantum public key distribution. It must be emphasized that a random phase for a reference QPKE is different from a random phase for a key QPKE, which adds extra difficulties to attackers without knowing the random phase for both

Glauber states. It is only Bob who can perform the calculation of Eq. (6) because only he precisely knows how to derandomize each of QPKEs. When Bob derandomizes a cipher QPKE, or decrypts a cipher QPKE, there is just a quantum operation to transform the cipher QPKE state without measurement, or no uncertainty occurs. That is a great advantage for Bob to take over the attacker.

Let's take a close look on this randomization process by introducing a phase shift operator or quantum phase shift gate $\hat{U}(\varphi) = e^{i\varphi}$ with a unitary relationship $\hat{U}^{\dagger}(\varphi)\hat{U}(\varphi) = 1$, for an arbitrary phase $\varphi$. Then we apply this unitary relationship to a Glauber state

$$|\alpha\rangle = \hat{U}^{\dagger}(\varphi)\hat{U}(\varphi)|\alpha\rangle = \hat{U}^{\dagger}(\varphi)|\alpha'\rangle$$

where $|\alpha'\rangle = |e^{i\varphi}\alpha\rangle$ as a randomized Glauber state to be transmitted by Bob over an optical channel to Alice. Bob keeps the reverse phase shift gate $\hat{U}^{\dagger}(\varphi)$ privately to be applied to the returning QPKE Glauber state before a coherent detection. From a quantum measurement point of view, $\hat{U}(\varphi)$ is an encoding basis of a Glauber state $|\alpha'\rangle$ and $\hat{U}^{\dagger}(\varphi)$ is the measurement basis of that Glauber state $|\alpha'\rangle$. A measurement without knowing the right measurement basis $\hat{U}^{\dagger}(\varphi)$ is a meaningless measurement. Here, Bob is the only one capable to make meaningful measurement by controlling a random basis per QPKE state.

When Bob applies his detector $\widehat{GSD}$ to perform an actual measurement on a QPKE, his result does contain measurement error in phase: $\pm\delta\varphi$. Using DPSK, Eq. (6) would have an error $\pm2\delta\varphi$. However, the modern coherent detections can adjust and remove the error through available techniques such as calibration based on a constellation diagram or a digital processing unit (DSP) and then obtain the correct key. Let's summarize the proposed quantum public key distribution in the following:

1. Bob generates a QPKE by applying the phase shift gate $\hat{U}(\varphi_r)$ with a secret phase $\varphi_r$, known only by him, and then transmits it over an optical channel to Alice.
2. Alice modulates a key phase $\varphi_k$, maybe also an amplitude, into the QPKE based on a random secret key and selected modulation scheme and returns it to Bob.
3. For the returning QPKE, Bob derandomizes it with his private key or the reverse phase shift gate $\hat{U}^{\dagger}(\varphi_r)$ and then measures it with a coherent detector to extract the secret by leveraging with DPSK.

For the proposed scheme, security considerations are important. In the following, we discuss different types of key modulations, and their ability to mitigate different existing major attacks on quantum communication systems. Additionally, we discuss how we optimize the randomization and modulation to achieve our best security goals.

*A. Key Modulation Schemes*

Optical coherent modulation and detection techniques have been well developed in the past decades and detail literatures can be found in a recent review paper [6]. The proposed QPKE method does support all existing modulation techniques. The most common modulations are:
- Quadrature phase-shift keying QPSK

  In this case, one QPKE state represents 2 bits of information. QPKE can be considered as a quantum system or qudit with 4 states.
- Amplitude phase-shift keying APSK

  This scheme modulates both phase and amplitude to a single pulse. APSK can work with a few modes such as 8-APSK for 3-bit modulations, 16-APSK for 4-bit modulations, 32-APSK for 5-bit modulations, 64-APSK for 6-bit modulations, etc.
- Phase-shift keying PSK

  PSK modulates phases only to signals such as 4-PSK with 4 modulation phases equivalent to QPSK, 8-PSK with 8 modulations phases, 16-PSK with 16 modulation phases.
- Differential phase-shift keying DPSK

  DPSK is widely used to avoid effects from communication channels. The encoding is based on the phase difference between the signal pulse and a reference pulse which is usually the direct previous pulse. However, DPSK can be extended beyond the direct previous pulse.

The modulation scheme must be selected based on supporting modulation types by the coherent detector $\widehat{GSD}$. In general, the proposal would work if both sender and receiver can communicate over an optical channel. The randomization module RM can be an independent unit in front of the detection unit. The detector unit can be an optical transceiver that transmits a signal pulse through RM, at which point RM's PM randomly modulates a phase into it and, when the pulse returns back, removes the random phase by its PM$^{\dagger}$ before sending the pulse back to the optical transceiver.

*B. Security Considerations*
- Man-in-the-middle (MITM) attack

  Some physical verifications for an optical channel can be used for authenticating the communication path. For example, Bob can authenticate Alice with optical fiber analysis tools, such as an optical time-domain reflectometer OTDR [13]. The physical testing and verification can help to maintain the optical channel's integrity. Within the context of the proposed QPKE approach, we can dynamically verify the integrity of an optical fiber path between communication peers by applying the signal time delay during the roundtrip based on the known optical fiber length. Detected time delay should be within the acceptable variation due to environment fluctuations. If the detected delay time is beyond the acceptable level, then action must be taken to identify the cause. The physical roundtrip channel with active runtime monitoring of the channel delay offers the capability to catch MITM attacks, disabling the requirement of a pre-shared secret for channel authentication as what we need in QKD.

- Intercept-resending attack

  Attackers can behave like Alice to completely intercept the QPKE signals and gain their intensities and phases, then regenerate them and send back to Bob. Due to the lack of signal-local oscillator LO's synchronization, plus the measurement errors from their detection equipment, it is physically impossible to generate QPKE signals with the same amplitudes and random phases as their values without interruptions, whereupon Bob can easily catch the attacks from the receiving bit error rate (BER). In this case, the no-cloning theorem can be also applied to an unknown Glauber state so the attacker can not exactly reproduce an intercepted Glauber state without causing higher BER at Bob.

- Tapping attack

  Generally, this attack is the best attack strategy for the proposed quantum public key distribution. The attacker must carefully make a small tapping to avoid a direct impact on the intensities of QPKE signals. Bob can monitor the receiving intensities and track to see if the variation is within the acceptable level. If not, action must be taken to identify the cause.

Let's take a closer look at tapping attacks. For an invisible weak tapping, the attacker must tap the same signal at two points T1 and T2 (see Fig. 4) and perform the measurements to remove the random phase $\varphi_r$ modulated by Bob. Although T1 should be close to Bob and T2 should be close to Alice to minimize the impacts to the QPKE states, the attacker may practically have to move the tapping point T1 close to Alice for possible operations. At T1, the attacker can gain

$$\Phi'_{T1} = \varphi_r + \varphi_{pT1} + \varphi_{LO-T1} \pm \delta\varphi, \qquad (7)$$

where $\delta\varphi$ in Eq. (7) is the measurement error from the attacker's detector and, at T2, the attacker can gain

$$\Phi'_{T2} = \varphi_k + \varphi_r + \varphi_{pT2} + \varphi_{LO-T2} \pm \delta\varphi, \qquad (8)$$

whereupon the attacker can apply Eq. (8) – Eq. (7) to remove the random phase $\varphi_r$,

$$\Phi'_k = \Phi'_{T2} - \Phi'_{T1}$$

$$= \varphi_k + \Delta\varphi_p + \Delta\varphi_{LO} \pm 2\delta\varphi \qquad (9)$$

Eq. (9) indicates that the random phase $\varphi_r$ is disappeared, but with the cost that three new terms appear: $\Delta\varphi_p$ from the path between T1 and T2, $\Delta\varphi_{LO}$ from the attacker's local oscillators, and the inevitable measuring errors $2\delta\varphi$ from equipment and the uncertainty principle in Eq. (2b). Comparing Eq. (9) with Eq. (6), the attacker has inevitable disadvantages from two measurements.

In order to remove $\Delta\varphi_p + \Delta\varphi_{LO}$, the attacker may apply the time delay from point T1 to point T2 to estimate $\Delta\varphi_p$ but the environment fluctuation may dramatically impact the accuracy of this estimation. Moreover, proper care must be taken to avoid the attacker's injecting her detection signal from T1 and splitting it out at T2. $\Delta\varphi_{LO}$ is much easier to remove if the attacker uses the same LO for both point measurements.

On the other hand, the uncertainty principle Eq. (2b) does not benefit an invisible tapping by the attacker because a smaller $\Delta n$ leads a bigger $\Delta\varphi$ when performing a measurement on a tapped pulse (see Figure 1). For a case of the average number of photons $\bar{n} = 10$ per coherent laser pulse, i.e. $\Delta n = \sqrt{\bar{n}} \approx 3$ photons, $\Delta\varphi \approx 0.016$ rad = $10°$. That means, the measurement on a tapped pulse containing about 10 photons renders an uncertainty of 10 degree in its phase. On the other hand, a commercially balanced 2x4 hybrid 90° couplers, used in the most of coherent detections, comes with an uncertainty of 5 degree directly from equipment. However, the improvement of detection equipment does not eliminate the contribution due to the uncertainty principle in Eq. (2b) for the invisible tapping. Without losing generality, we can take an overall uncertainty of 5 degree in a phase measurement as an example, and then Eq. (8) becomes

$$\Phi'_k \approx \varphi_k \pm 10 \text{ (degree)} \qquad (10)$$

Eq. (10) can be used as a basic guideline to select a right modulation scheme.

*C.* Optimizing Selection of Modulation

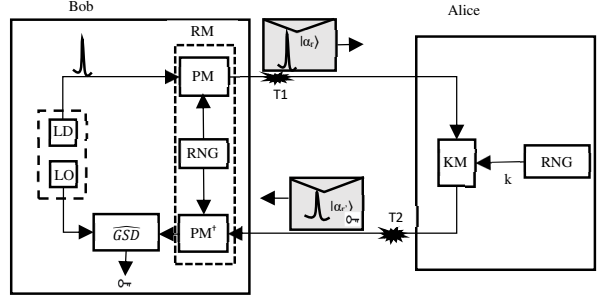

Figure 4. Possible tapping attacks are illustrated in this figure. Due to random phase modulated into a QPKE pulse, the attacker needs to perform tapping at point T1 and point T2 as shown in the diagram.

To achieve the optimal security from this proposal, we need to maximize those two advantages: randomization of QPKEs at Bob; and optimal key modulation at Alice. The randomization greatly increases the uncertainty on the attacker's measurements with a tapping technique, and the number-phase uncertainty principle would add an extra degree to her measurement errors; an optimal key modulation at Alice can practically render the attacker unable to distinguish one key phase from another, such that available digital processing techniques do not work for her:

- Randomization of Glauber states

  Use a good random number generator RNG and take the random number to decide the voltage for PM and PM$^\dagger$; due to self-controlled randomization, we can increase the number of random phases such as 1024 random phases to reduce the probability of guessing.

- Smart selections of key modulator KM

  Eq. (9) gives us a good guideline to take advantages from the attacker by selecting a modulator with a phase spacing less than 20 degree, that is, more than 16 phases. This renders the attacker unable to distinguish one key phase from another due to the physical limitation from a detection equipment and the uncertainty principle.

- Optimal choices of key modulator KM

  Differential phase-shift keying modulation or DPSK could be an optimal technique, in that Alice modulates her random key based on the phase difference with the reference QPKE in Eq. (6). From Eq. (9), the key phase difference for the attacker becomes

$$\Delta\varphi_k = \varphi_k - \varphi_R \pm 4\delta\varphi \qquad (11)$$

which means, DPSK doubles the uncertainty to the attacker's measurements due to a subtraction between key phase and reference phase. In this case, a modulation with 16 phases should be more than enough for security considerations.

- Dynamic phase reference list
  One more consideration could be useful to cause even greater difficulties for the attacker: allow the reference $\varphi_R$ to be dynamically chosen from a list of random reference phases based on a method agreed by both Bob and Alice. In this case, Eq. (10) does not work without knowing which reference is used. The reference list can be randomly generated at the beginning of the key distribution and updated periodically. The reference list can be considered as third advantage over the attacker and actively controlled by Alice. This allows a modulation with even fewer phases, such as QPSK, 4-PSK. However, if key phase spacing between neighboring phases is less than $4\delta\varphi$, the attacker is then practically unable to apply the tapping attacks due to a random distribution in her constellation diagram.

*D. Comparisons*

Unlike DV-QKD and CV-QKD where information carriers are considered as qubits, QPKEs are qudits with multiple states per carrier. QKD needs both quantum and classical channels from its protocol due to the uncertainty principle causing the uncertainties on measurements so the shared keys are established through the post-processing over the classical channel, together with a pre-shared secret to avoid MITM attacking. The proposed QPKE only requires single roundtrip optical channel with the self-controlled randomizing gate, mimicking RSA-type of key exchange mechanism. Key distribution is based on the matured optical communication techniques. The proposed protocol together with well considerations of the uncertainty principle and coherent state modulation schemes physically restricts the capability of attackers to apply potential attacks and achieves the security of key distributions. Thanks to the special characteristics of Glauber states from quantum at a low intensity to classical at a higher intensity, the proposal clearly reflects a central idea of the protocol design for highly secure key distributions: keeping the deterministic characteristic (classical) for communication peers and giving the maximum uncertainty (quantum) to attackers. In contrast to QKD where quantum repeaters are required for a longer distance, QPKE states can be amplified with mature technologies such as phase sensitive amplifiers [14] to reach a desired distance. In principle, QPKE can achieve a key rate at 50% speed of optical communication links, without special limitation of distribution distance from the protocol itself.

## IV. CONCLUSION

This paper proposes a novel quantum key distribution approach with a self-controlling randomized Glauber state as a quantum public key envelope, or QPKE. With a physical quantum system or Glauber state, it allows that the proposed system can be built with commercial optical modules widely available on the market today. The system runs over existing optical fiber networks. Possible optimizations of the proposed system with varieties of modulation schemes are explored. It is noted that DPSK with dynamic references would maximize the security of quantum public key distribution. In the future, we plan to implement this proposed system and demonstrate the key distribution over a telecom fiber optical channel.

ACKNOWLEDGMENTS

We want to acknowledge James Nguyen for his continued encouragement to explore public key exchange with physical quantum systems. We also want to express our appreciation to Prof. Rick Trebino from Georgia Institute of Technology, for intuitive discussions regarding measurements of weak laser pulses, as well as to Ken Dobell for his editorial assistance.